\documentclass[pre,twocolumn,groupedaddress]{revtex4}
\usepackage{graphicx}
\begin{document}
\title{A self-consistent approach to measure preferential attachment in networks and its application to an inherent structure network} 
\author{Claire P. Massen}
\affiliation{University Chemical Laboratory, Lensfield Road, Cambridge CB2 1EW, United Kingdom}
\author{Jonathan P.~K.~Doye}
\email{jpkd1@cam.ac.uk}
\affiliation{University Chemical Laboratory, Lensfield Road, Cambridge CB2 1EW, United Kingdom}
\date{\today}

\begin{abstract}

Preferential attachment is one possible way to obtain a scale-free network.
We develop a self-consistent method to determine whether preferential attachment occurs during the growth of a network, and to extract the preferential attachment rule using time-dependent data.
Model networks are grown with known preferential attachment rules to test the method, which is seen to be robust.
The method is then applied to a scale-free inherent structure network, which represents the connections between minima via transition states on a potential energy landscape.
Even though this network is static, we can examine the growth of the network as a function of a threshold energy (rather than time), where only those transition states with energies lower than the threshold energy contribute to the network.
For these networks we are able to detect the presence of preferential attachment,
and this helps to explain the ubiquity of funnels on energy landscapes. 
However, the scale-free degree distribution shows some differences from that of a model network grown using the obtained preferential attachment rules, implying that other factors are also important in the growth process.

\end{abstract}

\maketitle

\section{Introduction}

Scale-free networks are ubiquitous in nature \cite{Jeong00,Jeong01}, technology \cite{Albert99,Faloutsos99} and society \cite{Newman01c,Redner98}.
It is well known that a growth mechanism involving preferential attachment can 
lead to scale-free networks \cite{Barabasi99}.
In the original Barab\'asi-Albert (BA) model \cite{Barabasi99}, a new node is added at each time step together with $m$ new edges.
These new edges are more likely to attach to old nodes with high degree, thus the higher the degree of a node, the faster its degree increases, leading to the power-law degree distribution that is characteristic of scale-free networks.

For linear preferential attachment, nodes gain edges with a probability $\Pi$ that is proportional to their degree $k$.
For a more flexible power-law form, $\Pi \propto k^{\alpha}$, it has been shown that a scale-free network is only produced by the BA model if the exponent $\alpha=1$ \cite{Krapivsky00}.
For $\alpha<1$, the degree distribution follows a stretched exponential, and for $\alpha>1$ a single site connects to all the nodes.
However, in more complex models, the resulting networks can be scale-free even if $\alpha \ne 1$ \cite{Newman01b, Barabasi02, Jeong03, Eisenberg03, Roth05, Peltomaki05}.

Previously, we have analysed the properties of inherent structure (IS) networks \cite{Doye02,Doye05b,Massen05}, and 
here we want to test whether the preferential attachment model can help to explain the scale-free character of these networks.
IS networks represent potential energy landscapes by taking local minima in the potential energy, or inherent structures, to be nodes in the network, and edges link those minima that are directly connected by a transition state.
This network provides a dynamically relevant representation of the landscape, since the low-temperature dynamics of a system can be considered in terms of a hopping between basins of attraction surrounding the minima via transition states \cite{Stillinger84}.
IS networks for small Lennard-Jones clusters have been shown to be scale-free \cite{Doye02}. 
Furthermore, indirect evidence that this topology is more general
has come from the properties of the basins of attraction surrounding minima for supercooled liquids \cite{Massen05b}, 
and from the topology of more coarse-grained descriptions of the energy landscapes of proteins \cite{Rao04,Caflisch06}.

Understanding the origin of the topology of the IS networks is a particular challenge because, unlike many networks, they are static.
The topology is determined just by the interparticle potential and the size of the system.
Therefore, a preferential attachment approach might, at first sight, seem totally inappropriate to these networks because they do not grow in time.
However, here we wish to explore whether the IS networks can be understood in terms of a quasi-growth process.
In particular, we will examine how the network evolves as a function of a threshold energy, where only those transition states that lie below that energy give rise to edges in the network.
In this quasi-growth process the low-energy minima correspond to the `older' nodes in the network.
We have previously shown that there is a strong correlation between the energy of a minimum and its degree with the `older' lower-energy minima having higher degree \cite{Doye02}, thus suggesting that a preferential attachment model might be fruitful.

So far, preferential attachment has only been tested empirically in relatively few 
cases \cite{Newman01b, Barabasi02, Jeong03, Eisenberg03, Redner04, Roth05, Peltomaki05, Capocci06}.
This is partly due to the lack of time-resolved data for growing networks, but a reliable method for determining the form of the preferential attachment is also lacking.
The main problem is that the probability of a node with given degree gaining an edge, $\Pi(k,t) = k^{\alpha} / \sum_i k_i^{\alpha}$, depends on the size of the network through the denominator of this expression.

A method that has been used previously is to consider two snapshots of the network a certain time apart \cite{Barabasi02, Jeong03, Eisenberg03, Roth05}.
The relationship between the number of edges gained by a node in this time and its degree in the initial network can be determined.
The degree of each node, and hence $\sum_i k_i^{\alpha}$, 
is assumed to remain constant over this time period. The consequences of this approximation are not clear. Furthermore, 
the results will depend on the size of the time period used.
Actor, scientific collaboration and protein interaction networks have been studied using this method, finding exponents between 0.75 and 1.1.

Newman studied time-resolved data for scientific collaboration networks \cite{Newman01b}, comparing the probability that a node with given degree gains an edge (as measured from the data) with that expected if preferential attachment was not occurring.
This considers each step in sequence, but effectively assumes that the time-dependent constant, $\sum_i k_i^{\alpha}$, is proportional to the size of the network.
In the case that the average degree of the network remains constant and $\alpha=1$, this assumption is true.
However, the average degree of many networks increases \cite{Dorogovtsev01, Barabasi02, Mattick05, Gagen05}, and exponents different to one have been measured \cite{Barabasi02, Jeong03, Roth05}.

The above methods proved inadequate for analysing the quasi-growth of the 
IS networks. Therefore, to overcome the deficiencies of the above approaches, 
in Section \ref{sec:method} we introduce a self-consistent, iterative method to 
determine $\alpha$ from time-dependent data describing the growth of a network.
An initial trial value of the exponent $\alpha$ is assumed, and 
hence the constant of proportionality $\sum_i k_i^{\alpha}$ can be estimated, 
which in turn allows an improved estimate of the exponent to be obtained.
This process is repeated until convergence is achieved.
In Section \ref{sec:model}, the new self-consistent method is applied to model networks, 
for which the exponent for preferential attachment is known, in order to determine the accuracy of the method.
As the growth process is stochastic, some uncertainty in the value of $\alpha$ is inevitable.
If each node gained precisely $k^{\alpha}/\sum_i k_i^{\alpha}$ edges at each time step, the method would be exact.
Finally, in Section \ref{sec:lj13}, 
we apply this method to the IS network for the 13-atom Lennard-Jones cluster, LJ$_{13}$.

\section{Method}
\label{sec:method}

When a new node is added to the network at time $t$, we assume that the degree of the existing nodes to which the new node connects are chosen with probability $\Pi(k,t) \propto n_k(k,t) f(k)$, where 
$n_k(k,t)$ is the number of nodes with degree $k$ at time $t$, and 
we assume $f(k)$ to be time independent and 
to have the preferential attachment form $f(k)=k^{\alpha}$. 
In one time step, the expected number of edges gained by all nodes with degree 
$k$ is then
\begin{equation}
\Delta k (k,t) = m(t) \Pi (k,t) = \frac {m(t) n_k(k,t) f(k)} {c_t(t)},
\label{eqtn:f}
\end{equation}
\noindent where $m(t)$ is the number of new edges added and $c_t(t)=\sum_{k'} n_k(k',t) f(k')$.

The basic idea for determining $f(k)$ is based on inverting the above expression.
$m$, $\Delta k$ and $n_k$ can be simply measured at each time step, 
and $c_t$ can be estimated by using a trial exponent.
A fit to $f(k)=k^{\alpha}$ gives a new exponent, which is then used to improve the estimate of $c_t$.
This process is repeated until self-consistency is achieved, i.e.~the exponent converges.

More specifically, rearranging Eq.~\ref{eqtn:f} gives
\begin{equation}
f(k) = \frac {\Delta k(k,t) c_t(t)} {m(t) n_k(k,t)},
\label{eqtn:f2}
\end{equation}
As $f(k)$ is assumed to be time independent, this could then be averaged over all 
appropriate time steps.
However, Eq.~\ref{eqtn:f2} diverges if $n_k(k,t)=0$.
Therefore, to avoid this situation we sum both sides of Eq.~\ref{eqtn:f} over $t$, effectively comparing the number of edges gained by all nodes with degree $k$ at all times, $\sum_t \Delta k (k,t)$, with the expected number, $\sum_t (m(t) n_k(k,t) f(k)/c_t(t))$.
This then gives
\begin{equation}
f(k) = \frac {\sum_t \Delta k (k,t)} {\sum_t (m(t) n_k(k,t)/c_t(t))}. 
\label{eqtn:b} 
\end{equation}
If one wished, the assumption that $f(k)$ is independent of time could be checked 
by calculating $\alpha$ for different time ranges.

Finally, $\alpha$ is obtained from $f(k)$ using a linear least squares fit for $\log f=\alpha \log k$.
However, the statistics for high-degree nodes are poor, as there are less of them and they only occur at later time steps.
There are high-degree nodes that are not chosen to get any new edges, giving $f(k) = 0$ which cannot be taken into account in the log-log fit.
An approach that has been used previously is to plot the cumulative distribution, giving an exponent of $\alpha-1$ \cite{Barabasi02, Jeong03, Roth05}.
However, for the examples we consider, we find that logarithmic binning over degree gives better results.
The contribution of each degree to the bin is weighted by the number of time steps at which there are nodes with that degree, i.e.~the number of terms in the sums of Eq.~\ref{eqtn:b}.
This weighting was also used in the least squares fit.

Many networks also gain new edges between two existing nodes.
These have been termed `internal edges', whilst new edges connecting to a new node are `external edges' \cite{Albert00b}.
When internal edges are added, the average degree of the network increases with time and the network is said to be accelerating \cite{Dorogovtsev01, Mattick05, Gagen05}.
This increases the global connectivity of the network, and is important in regulatory networks, such as biological networks \cite{Mattick05}.
The current method for measuring preferential attachment can equally well be applied to internal links, where the product of the degrees of the pair of existing nodes is chosen with probability $\Pi (k_ik_j,t) \propto n_k(k_ik_j,t)f(k_ik_j)$, and $f(k_ik_j)=(k_ik_j)^{\alpha}$.
Thus, preferential attachment can be studied separately for internal and external links.

\section{Model Networks}
\label{sec:model}

To test the method described in the previous section, we first apply it to 
model networks that have been grown using preferential attachment 
to see if the method can correctly recover the exponent used.
As we are developing this method in order to apply it to an IS network,
we choose the model networks to have some of the basic properties as the IS network
that we consider in Section \ref{sec:lj13}, 
e.g.\ the number of nodes (1509) and edges (20687). 
Furthermore, at each time step in the growth of the networks, 
one of three possible processes occurs, namely the addition of external or internal edges, or the addition of two new nodes with a edge between them, and
the relative frequencies of the three events, which have a strong effect on network 
topology \cite{Albert00b, Dorogovtsev01}, were taken to be the same as for 
the quasi-growth of the IS network.
One edge is added at each time step, 
giving $m=1$ for all times, also as for the IS network.
New links between two new nodes cannot preferentially attach, but they are necessary to reproduce the correct event frequencies, and to avoid complications from initial conditions.

A further complication associated with the introduction of internal links is that this can lead to multiple edges (MEs) and self-connections (SCs), which can have a significant effect on network topology \cite{Maslov04}.
In the case of external links, ME/SCs are only relevant if more than one edge is added with the new node.
ME/SCs are not present in the IS network, as is true of many other networks, and so they are excluded from the model networks.
To exclude ME/SCs when adding an internal link, two nodes are chosen from the set of pairs of nodes that would not give an ME/SC.
This set of pairs are also used to determine $c_t$ for these time steps.

\begin{figure}
\centerline{\includegraphics[width=8.6cm]{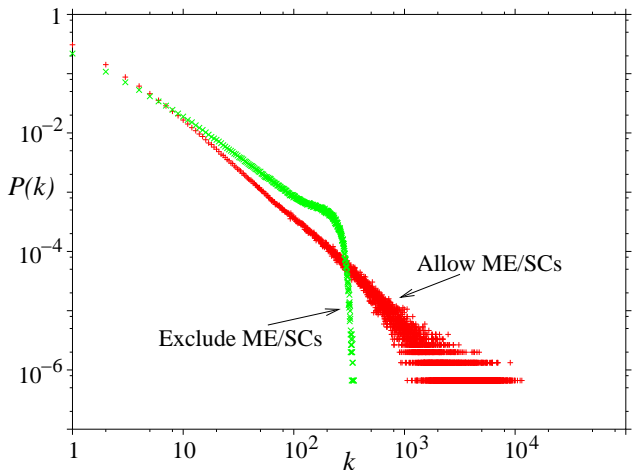}}
\caption{(Colour online) Degree distributions averaged over 1000 networks, with 1509 nodes, for versions of the model excluding and allowing ME/SCs.}
\label{fig:pk}
\end{figure}

In the final network, there will be less edges between high-degree nodes compared to a network where ME/SCs are allowed.
When a new edge is added, high-degree nodes are likely to be connected already and so cannot be connected if ME/SCs are excluded, but would be likely to be chosen if ME/SCs are allowed.
In terms of the degree distribution, the maximum degree will be lower.
The degree distribution obtained from this model is shown in Fig.~\ref{fig:pk}.
If ME/SCs are excluded, there is a cut-off at high degree which is not present when ME/SCs are allowed.
The cut-off is particularly extreme in this case as the network is very dense with high average degree ($ \langle k \rangle = 27.4$).

The self-consistent method described is very successful in determining the preferential attachment rules in the model network.
It is most accurate when the exponent is close to one (Table \ref{table:trials}),
e.g.\ when $\alpha=1$ we obtain a result of 0.988 for external edges and 0.996 for internal edges, independent of the trial exponent used.
Reasonable results are obtained for $0.5<\alpha<1.5$, with a maximum error of around 0.12, increasing as $\alpha$ differs more from 1.
Exponents very different to one are unlikely to give scale-free networks \cite{Krapivsky00}, so it is reasonable to expect $\alpha$ to lie within this range.

\begin{figure}
\centerline{\includegraphics[width=8.6cm]{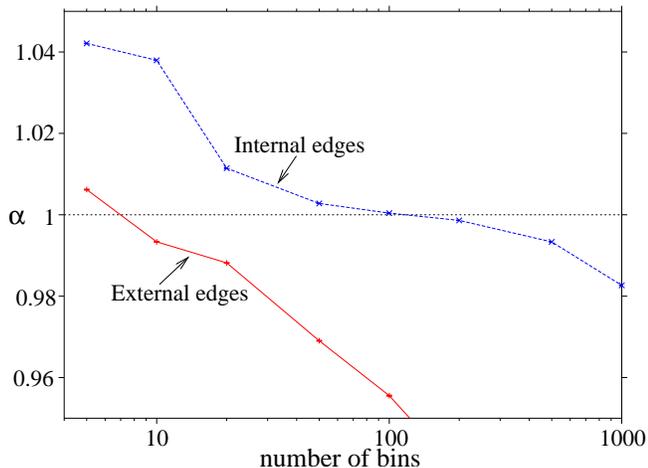}}
\caption{(Colour online)
Variation of the exponent with the number of bins used in the least squares fit for the model, for (a) external and (b) internal links.
The exponent used to make the network, $\alpha=1$, is also plotted.
Error bars are standard error from the linear least squares fit.}
\label{fig:bins}
\end{figure}

The exponent is somewhat sensitive to the the number of bins used in the least 
squares fit, as shown in Fig. \ref{fig:bins}, but 
the result is within 0.05 of the input exponent for reasonable choices for the
numbers of bins.
For very few bins, the least squares fit is inaccurate, whereas
for very many bins, a significant number have zero value and 
so will not contribute.
For both internal and external links, the slope is flatter between these extremes, implying that the method is robust with respect to the number of bins in that range.

\begin{table}
\caption{\label{table:trials}
Model networks were grown using a given input exponent, and the output exponent was measured for external and internal edges.
Results are for one realisation of the model network for each input exponent.
}
\begin{ruledtabular}
\begin{tabular}{ccc}
Input exponent   & \multicolumn{2}{c}{Output exponent} \\ 
\cline{2-3}
                 & external edges & internal edges \\
\hline
0.5              & 0.4817  & 0.4807  \\
0.8              & 0.7823  & 0.8053  \\
1.0              & 0.9882  & 0.9955  \\
1.2              & 1.2038  & 1.2044  \\
1.5              & 1.6245  & 1.5594  \\
\end{tabular}
\end{ruledtabular}
\end{table}

The optimal number of bins for both internal and external links is consistent 
with approximately 60 data points per bin.
The actual number of bins reflects the relative frequencies of the two types of 
link; for the current examples there are 19205 internal links and 
1455 external links.
Analysis of 100 networks revealed that for external links, 
25 bins is the optimal number, giving a mean $\alpha$ of 1.00067 and 
a standard deviation of 0.02280.
Note that the method gives different results on different realisations of the network because of the stochastic nature of the network growth.
For internal links, the optimal number of bins was 300, 
with mean 1.00067 and standard deviation 0.00593.

\begin{figure}
\centerline{\includegraphics[width=8.6cm]{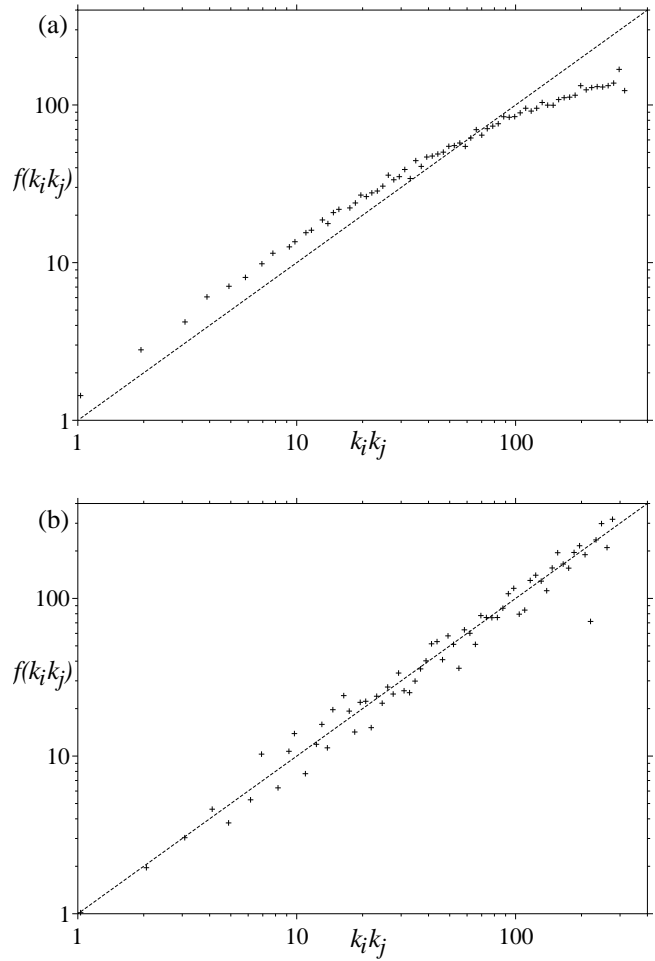}}
\caption{
$f(k_ik_j)$ against $k_ik_j$ for internal links when (a) ME/SCs are excluded in making the network, but allowed in the analysis,
(b) ME/SCs are also excluded in the analysis.
$\alpha=1$ was used in the growth of the network, this is also shown.
The trial exponent is 1.
}
\label{fig:fkikj}
\end{figure}

The method would be simpler if ME/SCs could be ignored when adding internal links.
To test this, ME/SCs were excluded while making the network, but allowed in the analysis.
This gives $\alpha=0.779$ for internal links, with an input exponent of one.
This should be compared to $\alpha=0.996$, obtained when ME/SCs were excluded in the analysis.
The cause of the problems can be understood from Fig.~\ref{fig:fkikj}(a),
which shows $f(k_ik_j)$ against $k_ik_j$.
$f(k_ik_j)$ is too low for pairs of nodes with high $k_ik_j$, 
and too high for pairs with low $k_ik_j$.
There are less edges between high-degree nodes than the analysis expects, because ME/SCs are excluded and high-degree nodes are likely to be connected already.
Therefore, there are more edges between low-degree nodes.
For comparison, $f(k_ik_j)$ is shown in Fig.~\ref{fig:fkikj}(b) for internal links with ME/SCs excluded in both the creation of the network and its analysis, giving a straight line with exponent close to 1.

We also tested some of the alternative methods for extracting 
$\alpha$
that were described in the 
introduction \cite{Newman01b, Barabasi02, Jeong03, Eisenberg03, Redner04, Roth05, Peltomaki05, Capocci06}. 
We just give results for the current model networks, when the input exponent
is one for both internal and external links.
Using the method that follows the number of edges obtained over a certain time 
interval by nodes with a given degree at the start of that interval \cite{Barabasi02, Jeong03, Eisenberg03, Redner04, Roth05}, we obtain $\alpha \approx 0.7$. 
The method of Newman \cite{Newman01b, Peltomaki05, Capocci06}, which approximates the time-dependent constant in $\Pi(k,t)$, 
gives an exponent of $\alpha \approx 0.94$ for external edges, but 
for internal edges, a least-squares fit gives an exponent of approximately 0.4.
The plot is close to linear at low degree, but has a cut-off at high degree similar to that of Figure \ref{fig:fkikj}, which was due to the effects of ME/SCs.

That the performance of these alternative methods on the current networks are 
significantly worse than our self-consistent approach is unsurprising
given the greater number of approximations involved.
Furthermore, it is also not straightforward to account for effects due to 
the exclusion of ME/SCs in either of these alternate methods, 
leading to further inaccuracies.
However, we should also note that the current examples represent somewhat atypical
networks. In particular, they are unusually dense with a higher average degree than many other networks. 
Moreover, our aim was not to attempt to exhaustively compare the different
methods of detecting preferential attachment, 
but rather only to determine which is most appropriate to study the IS networks.

\section{Connecting up an Energy Landscape}
\label{sec:lj13}

In previous analyses of IS networks, each transition state gives rise to an edge in the network (except when it leads to an ME/SC) no matter what the energy of the transition state involved.
Here, we want to investigate the relationship between the IS networks and the topography of their potential energy landscape with a view to better understanding the scale-free character of the IS networks.

\begin{figure}
\centerline{\includegraphics[width=8.6cm]{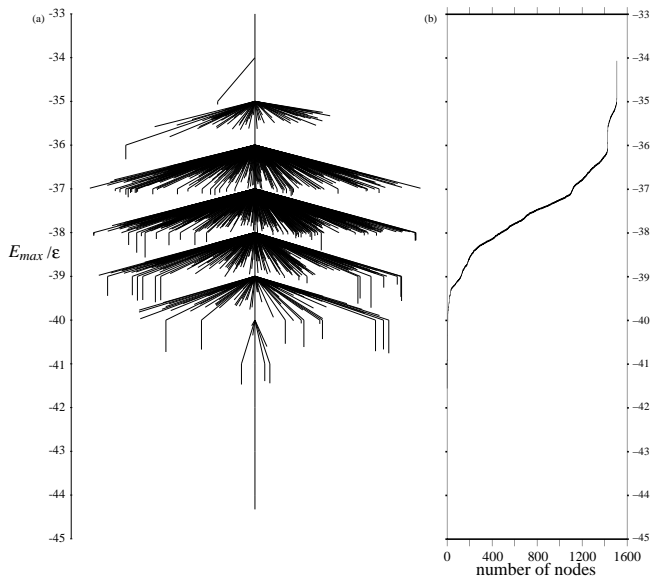}}
\caption{(a) Disconnectivity graph for LJ$_{13}$.
As the maximum transition state energy, $E_{max}$, is increased, more minima become interconnected.
(b) The increase in the number of nodes in the IS network with $E_{max}$.
The unit of energy is $\epsilon$, the pair well depth of the Lennard-Jones potential.}
\label{fig:13discon}
\end{figure}

One method that has been particularly successful at visualising the topography of multi-dimensional potential energy landscapes is the disconnectivity graph \cite{Becker97, Wales98, Doye99b, Landscapes}.
The disconnectivity graph for LJ$_{13}$ is illustrated in Fig.~\ref{fig:13discon}(a).
Each line in the graph represents a set of minima that are connected by paths that lie below a given energy threshold, $E_{max}$.
Each line begins at the energy of the corresponding minimum.
As $E_{max}$ increases transition states become accessible that link up sets of minima that were previously disconnected, and the corresponding lines in the disconnectivity graph join up.
Finally, at large $E_{max}$ the network has a giant component that spans the whole potential energy landscape and there is a single line in the disconnectivity graph.
When viewing Fig.~\ref{fig:13discon}(a), one should note that the disconnectivity analysis has been performed at a series of discrete energy levels.
This is purely for convenience in the generation and visualization of these graphs.

Disconnectivity graphs are useful because they provide a visual representation of the barriers between different regions of the energy landscape.
In the case of LJ$_{13}$, the landscape has a single funnel topography, and this is clear from Fig.~\ref{fig:13discon}, which has a single dominant stem, i.e.~most minima join directly to the set of minima containing the global minimum, and the barriers involved are relatively small.

Here, in a similar way, we wish to examine the topology of the IS network, as $E_{max}$ increases, where only those transition states with energy less than $E_{max}$ can contribute edges to the network.
Thus, we can look at the dynamics of growth of the network, with $E_{max}$ playing a similar role to time.
As the network grows one would expect new nodes and edges to be more likely to connect to lower-energy minima that have larger basins of attraction and larger degree.
Indeed, for the complete network a strong correlation between degree and minimum energy has been noted \cite{Doye02,Doye05b}.
Therefore, we decided to investigate whether this growth of the network might be describable in terms of a preferential attachment model.

$E_{max}$ is increased such that only one transition state is added at each step, and minima (nodes) are only included in the network once they become connected to another minimum by a transition state.
Edges in the network correspond to transition states, but not all transition states lead to edges, as some connect permutational isomers of the same minimum, forming self-connections, and more than one transition state may connect the same pair of nodes, forming a multiple edge.
Transition states that would form ME/SCs are ignored.

\begin{figure}
\centerline{\includegraphics[width=8.6cm]{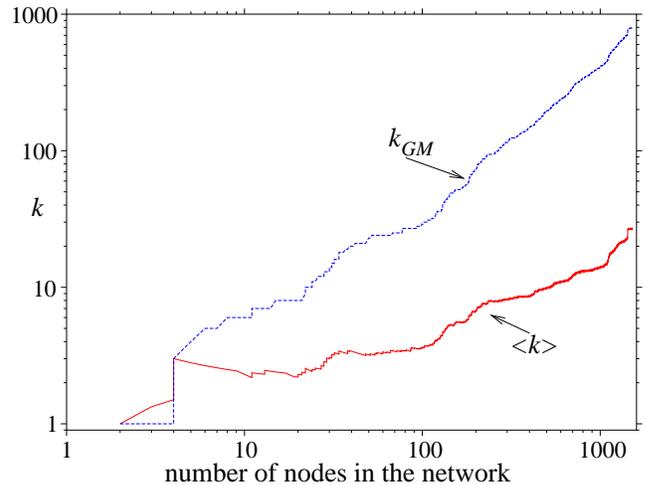}}
\caption{(Colour online)
Average degree, $\langle k \rangle$, and degree of the global minimum, $k_{GM}$, for the IS network as a function of nodes in the network.
}
\label{fig:kav}
\end{figure}

Fig.\ref{fig:13discon}(b) illustrates how the number of nodes in the network increases with $E_{max}$.
The distribution of the number of minima as a function of energy is expected to be a Gaussian for bulk systems \cite{Sciortino99, Buchner99, Heuer00}, and even for a small cluster such as LJ$_{13}$, the distribution is a reasonable approximation to a Gaussian \cite{Doye05b}.
Therefore, the rate of growth of the number of nodes in the IS network goes through a maximum at an $E_{max}$ value near to the centre of this distribution (it will actually be slightly above, because a minimum is only added to the network once it becomes connected to another minimum by a transition state with energy less than $E_{max}$).

The number of edges in the network increases faster than the number of nodes, as $M \sim N^{1.56}$.
The distribution of transition states as a function of energy is also expected to be Gaussian, but where the centre of the Gaussian is displaced to higher energy compared to minima \cite{Doye02b,Shell04}.
Therefore, the number of edges increases more rapidly than the number of minima at larger $E_{max}$.
This result implies that the average degree increases as $\langle k \rangle \sim N^{0.56}$, as shown in Fig.~\ref{fig:kav}, i.e.~the network is accelerating.

\begin{figure}
\centerline{\includegraphics[width=8.6cm]{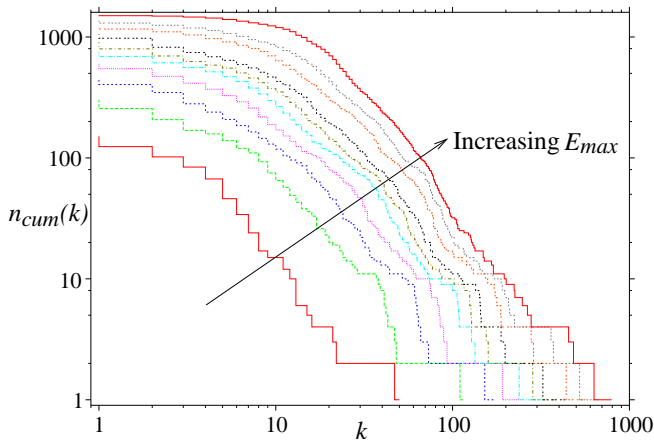}}
\caption{(Colour online)
Cumulative degree distribution for the IS network at different stages of the quasi-growth process.
}
\label{fig:pkn}
\end{figure}

From Fig.~\ref{fig:pkn}, we can see how the degree distribution develops during the quasi-growth process.
The overall form of the distribution is quite similar throughout the growth process, although the scale-free tail becomes more pronounced as the network grows and fluctuations are reduced.

These features affect other network properties.
For example, $M$ increases more slowly than $N^2$, so the network becomes more sparse.
This causes the clustering coefficient to decrease, although the network becomes more highly clustered than a random network with the same degree distribution, i.e.~the clustering becomes more significant.
Secondly, the accelerating nature of the network affects the average shortest path length of the network, $l_{ave}$.
If the random graph result for $l_{ave}$ were to apply, i.e.~$l_{ave} \sim \log N / \log \langle k \rangle$, then $l_{ave}$ for an accelerating graph should remain constant during growth \cite{Dorogovtsev03}.
In fact, after an initial increase the average shortest path length actually decreases with further growth.

Indicators of preferential attachment can be found in the evolution of some of the network properties.
For example, Fig.~\ref{fig:kav} shows the increase in degree of the global minimum, the `oldest' node, and the one which has the highest degree in the final network, as the network grows; it has the form $k_{GM} \sim N^{0.98}$.
The exponent describing how the degree of a node increases with network size (number of nodes) can be measured for each node, and is distributed about one, implying that this behavior of the global minimum is fairly typical.
The degree of a particular node is expected to grow faster than the average degree, because the addition of new nodes with $k=1$ reduces the average degree.
However, $k_{GM}$ increases much faster than expected from this effect, indicating that preferential attachment is significant during growth.

Applying the self-consistent method of Section \ref{sec:method}, the exponents for preferential attachment were determined to be $\alpha=0.87 \pm 0.05$ for external links, and $\alpha=0.63 \pm 0.05$ for internal links.
These exponents depend on the number of bins used in the least squares fit, as did those for the model network.
However, the variation is very similar to that for the model networks in Section \ref{sec:model}, again allowing an appropriate number of bins to be chosen.

\begin{figure}
\centerline{\includegraphics[width=8.6cm]{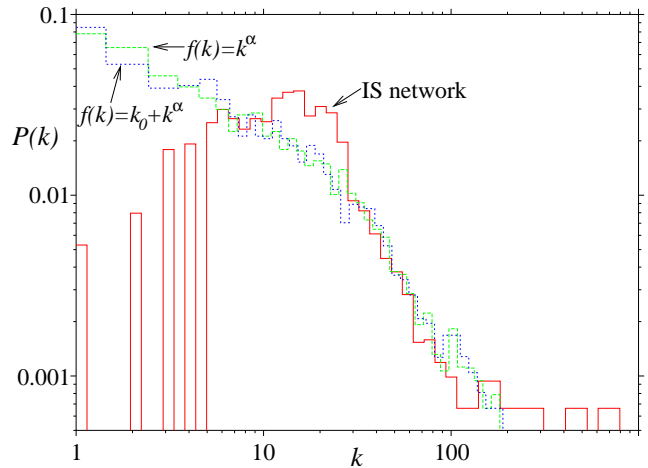}}
\caption{(Colour online) Degree distributions for the IS network and two model networks using $f(k)$ taken from the IS network results.
The first model uses $f(k)=k^{0.87}$ and $f(k_ik_j)=(k_ik_j)^{0.63}$, 
the second uses $f(k)=67.53+k^{1.56}$ and $f(k_ik_j)=(k_ik_j)^{0.63}$ for external and internal links respectively.
For clarity, results are shown for one realisation of the model networks.
When averaged over 1000 networks the results are qualitatively the same.
}
\label{fig:pk2}
\end{figure}

Fig.~\ref{fig:pk2} shows the degree distribution obtained for the model using the exponents found for the IS network.
Note that the degree distribution for the IS network is only a power-law over a finite range.
The degree distribution for the model is very similar to that for the IS network for mid-degree nodes, $30 < k < 100$, in the range where the IS network is scale-free.
Perhaps unsurprisingly, the model fails in reproducing the very high-degree nodes with $k > 200$, and the behavior at low degree, where the IS network deviates from a scale-free distribution.

\begin{figure}
\centerline{\includegraphics[width=8.6cm]{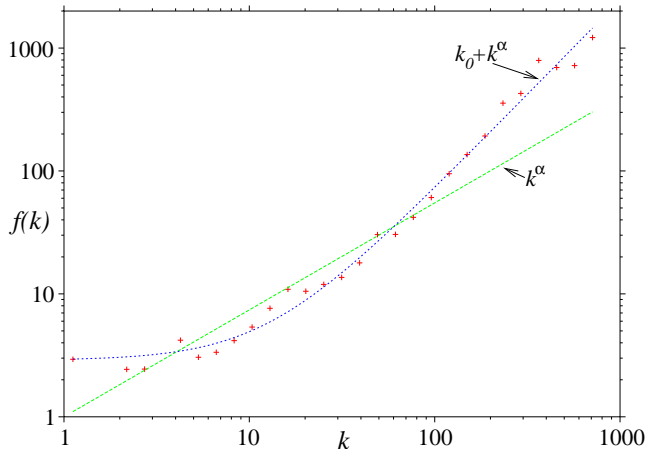}}
\caption{(Colour online)
$f(k)$ against $k$ for the IS network after convergence of our self-consistent approach to $\alpha=0.87$, assuming $f(k)=k^{\alpha}$. 
The best fits for $f(k)=k^{\alpha}$ and $f(k) \propto k_0+k^{\alpha}$ are also shown.
}
\label{fig:fk}
\end{figure}

In our analysis, correlations between nodes are not considered.
However, the IS network is assortative with respect to energy \cite{Doye05b}, i.e.~nodes with similar energies are more likely to be connected.
When a new node appears and creates a new external link, the nodes that have the most similar energies are the newest nodes, with lowest degrees.
Therefore, $f(k)$ for low-degree nodes is higher than expected for a preferential attachment rule, as can be seen in Fig.~\ref{fig:fk}.
The method used so far relies on $f(k)$ following a power-law, but can be adapted to use different functions for $f(k)$.
The only difficulty arises in extracting coefficients from the data for $f(k)$.
As a rough estimate, a power law does describe the slope fairly well.
In this case, a better fit is provided by $f(k)=k_0+k^{\alpha}$.
Using this form for $f(k)$ in our self-consistent approach gives $f(k) = 67.33+k^{1.56}$.
Compared to the original power-law fit, the exponent is now much greater, showing strong preferential attachment and increasing the number of external edges gained by high-degree nodes.
Large $k_0$ means that low-degree nodes gain more external edges.
This leads to the degree distribution shown in Fig.~\ref{fig:pk2}.
It is clear that improving the form of $f(k)$ has only a minor effect on the degree distribution.
The power-law form, which is easier to implement, is a good approximation, but care should be taken in attaching significance to the value of the exponent.

\section{Conclusion}

We have introduced a self-consistent method to measure preferential attachment in any network where time-resolved data is available.
Tests of the method on model networks where the form of preferential attachment is known show that the method can determine this preferential attachment accurately, although any such method will involve some uncertainty as network growth is a stochastic process.
The method is not limited to power-law preferential attachment rules.
However, this can provide a simple approximation to more complex forms, giving similar final degree distributions.
Other topological properties, such as assortativity, may be affected by this approximation.
We have also shown that multiple edges and self-connections can be very important, particularly if the networks studied are dense.
Excluding ME/SCs leads to a cut-off at high degree in the degree distribution of the resulting network, and failure to take the absence of ME/SCs into account 
when measuring preferential attachment underestimates its strength.

We have then applied this method to an IS network, which has a scale-free degree distribution.
Although such networks are static, a helpful way to think about the properties of the landscape is in terms of how the landscape becomes connected up as a threshold energy $E_{max}$ is raised.
We were able to show that during this quasi-growth process, there is preferential attachment, i.e.~that as new minima are added to the network they are more likely to add to the lower-energy high-degree nodes.
Therefore, the topography of the energy landscape plays a key role in the development of the scale-free topology of the IS network.

This interplay of topology and topography can help us understand why features 
like funnels \cite{Leopold,Bryngelson95} are so generic to energy landscapes.
At the bottom of such a funnel is a deep energy minimum that has no further 
downhill connections.
The preferential attachment implies that higher-energy minima are likely to be 
connected to lower-energy minima, which, if not themselves funnel bottoms, 
are in turn are likely to be connected to even lower-energy minima, and so on. 
This naturally generates features on the energy landscape, 
where there is a convergence of pathways down to a particular low-energy 
structure, and where there is an overall gradient of the landscape down 
towards this structure. These are the archetypal characteristics of a funnel on 
an energy landscape, a landscape object that has featured so 
prominently in the understanding of how relaxation to a target low-energy 
configuration is possible on a high-dimensional energy landscape, 
such as in protein folding \cite{Dobson98}.

On a relatively low-dimensional landscape, such as for LJ$_{13}$, 
the landscape is dominated by a single funnel surrounding the global minimum. 
But, this does not have to be the case for higher-dimensional landscapes, and
so when there are deep minima that are well-separated in configuration space, 
each can give rise to their own funnel, thus resulting in a multiple-funnel
landscape \cite{Doye99,Doye99b}. 
On a purely topological level, and in network parlance, these
funnels are evident in terms of different communities surrounding 
high-degree nodes \cite{Massen05}.

A model network grown using the preferential attachment rules 
discovered for the IS network reproduced the scale-free part of the 
degree distribution of the IS network fairly well.
However, there were discrepancies at high and low degree, indicating that preferential attachment alone is not sufficient to describe the whole degree distribution.
Of course, we could get an increasingly accurate description of the development
of the network by using increasingly complex forms for $f(k)$. However, as the forms
become more complex, they are also likely to produce less physical insight. The
use of a two-parameter expression for $f(k)$ did, though, 
confirm the robustness of the preferential attachment result, and illustrated 
that the exact value of the exponent should be interpreted with caution. 

It is also possible to consider how the IS network changes as the form of the potential describing the interatomic interactions in the cluster is changed.
In particular, as the range of the potential decreases, new minima and transition states appear on the potential energy landscape \cite{Doye96,Doye96c,Wales01}.
This will be studied in future work.

\begin{acknowledgments}
The authors are grateful to the Engineering and Physical Sciences Research Council
(CPM) and the Royal Society (JPKD) for financial support.
\end{acknowledgments}

\end{document}